\documentclass[preprint,aps]{revtex4-1}
\usepackage{graphicx}

\usepackage{natbib}
\usepackage{wrapfig}

\usepackage{graphics}

\usepackage{natbib}

\newcommand{\be}{\begin{equation}}
\newcommand{\ee}{\end{equation}}
\newcommand{\nn}{\mbox{} \nonumber \\ \mbox{} }
\newcommand{\ba}{\begin{eqnarray}}
\newcommand{\ea}{\end{eqnarray}}

\newcommand{\Alfven}{ Alfv\'{e}n }

\newcommand{\E}{{\vec{E}}}
\newcommand{\B}{{\vec{B}}}

\newcommand\eg{\textit{e.g.,\ }}
\newcommand\cf{\textit{cf.\ }}

\newcommand{\Bf}{{magnetic field}}
\newcommand{\Bfs}{{magnetic fields}}
\newcommand{\Ef}{{electric  field}}

\newcommand{\NS}{neutron star}
\newcommand{\NSs}{{neutron stars}}
\newcommand{\EM}{electromagnetic}
\newcommand{\BH}{{black hole}}
\newcommand{\BHs}{{black holes}}
\newcommand{\Sc}{Schwarzschild}
\newcommand{\ms}{magnetosphere}
\newcommand{\mss}{magnetospheres}

\begin{document}

\title{Electromagnetic power of  merging and collapsing compact objects}
\author{Maxim Lyutikov\\
Department of Physics, Purdue University, \\
 525 Northwestern Avenue,
West Lafayette, IN
47907-2036 }

\begin{abstract}
Understanding possible  \EM\ signatures of the merging and collapsing compact object is important for identifying possible sources of LIGO signal. Electromagnetic emission can be produced as a precursor to the merger, as a prompt emission during the collapse of a \NS\ and  at the spin-down stage of the resulting  Kerr-Newman \BH.
For  the NS-NS mergers,  the precursor power scales as  $L \approx B_{\rm NS}^2 G  M _{\rm NS} R_{\rm NS}^8/ ( R_{\rm orb}^7 c)$, 
% where $B_{\rm NS}$ is the \NS's \Bf, $M_{\rm NS} $ is its mass and $R_{\rm orb}$ is the radius of the orbit. 
while for the  NS-BH mergers,   it is $ ( G  M /( c^2 R_{\rm NS}))^2$ times smaller. 

We demonstrate that the
 time evolution of the axisymmetric force-free magnetic  fields can be expressed in terms of  the 
 hyperbolic   Grad-Shafranov equation and  formulate the generalization of the   Ferraro's law of iso-rotation to time-dependent angular velocity.
We  find exact non-linear  time-dependent  
 Michel-type  (split-monopole) structure of \mss\  driven by spinning and collapsing \NS\ in \Sc\ geometry. 
 % Michel's solution of the  Grad-Shafranov equation remains valid for arbitrary  time-dependent angular velocity of the  foot-points. 
 
 Based on this solution,  
 we argue that    the collapse of a  NS into the  BH happens smoothly, without  natural formation of current sheets or other dissipative structures on the open field lines and, thus,  does not allow the \Bf\ to become disconnected from the  star and  escape to infinity.  Thus,  as long as an  isolated Kerr   \BH\ can produce plasma and currents, it does not lose its  open \Bf\ lines, its magnetospheric structure evolved towards a split monopole  and the \BH\  spins down electromagnetically (the closed field lines get absorbed by the hole). The "no hair theorem", which assumes that the outside medium is a vacuum,  is not applicable in this case:  highly  conducting plasma introduces a topological constraint  forbidding the disconnection of the    \Bf\ lines from the \BH. Eventually, a single random  large scale spontaneous reconnection event   will lead to magnetic field release,  shutting down the \EM\  \BH\ engine forever. 
  Overall, the \EM\ power in all the above cases is expected to be relatively small.
 
We also  discuss the nature of short Gamma Ray Bursts and  suggest  if the \Bf\ is amplified to $\sim 10^{14} $ G during the merger  or the core collapse, the similarity of  the  early afterglows   properties of long and short GRBs can be related to the fact that in both cases a spinning  \BH\ can retains \Bf\ for sufficiently long time to extract a large fraction of its  rotation energy and produce high energy emission via the internal dissipation in the wind.  
 \end{abstract}

 \maketitle

\section{Introduction}

Estimating  possible \EM\ signature of merging and collapsing \NSs\ is most desirable for the gravitation waves searchers by LIGO and for  identifying possible    progenitors of short Gamma Ray Bursts. 
Collapse of a \NS\ into \BH\ may proceed either through the accretion induced collapse  (AIC) or during  binary \NS\ mergers. We expect at late stages  both processes proceed along a somewhat similar path: in case of the merger,  the two collapsing  \NSs\ form a transient  supermassive \NS\ which  then collapses into the \BH. 
Both  an accreting \NS\ (in case of an AIC) and  the transient supermassive \NS\ are expected to be magnetized. In addition, 
in case of merging \NSs\ the strong shearing of the matter may increase \Bf\ well above the initial values. 

In case merger of compact stars the \EM\ power can be generated as a precursor to the merger  due to either effective friction of the \NS\ \mss, or due to purely general relativistic effect, 
see \S  \ref{precur}
Later, and in the case of the AIC, 
 several types of \EM\ emission can be foreseen. First, the  \EM\  power in vacuum may be generated directly, due to the changing magnetic moment of the collapsing star \citep{GinzburgOzernoi,2003ApJ...585..930B}. Even if the  outside medium is highly conducting, \EM\ may be generated  
 via effective (resistive) disconnection of the external \Bfs, provided that the collapse naturally leads to formation of narrow dissipative current structure. 
Second,  a  pulsar-like \EM\ power generated by the rotation of the {\NS}s and extracted via the magnetic field.  As we argue below, as long as the \BH\ can produce plasma via vacuum breakdown, it can self-generate electric currents,   retain the \Bfs\ and spin-down electromagnetically for time periods much longer than the collapse time, see   \S 
\ref{GS-time}.

Conventionally, in estimating the possible  \EM\ signatures it was  first assumed that a fraction  $ {R_{\rm NS}/ R_G}$ of the initial external magnetic energy (also built-up by the collapse and compression of the  \Bf)
is radiated away on time scale of the order of the collapse time \citep{1975ARA&A..13..381E}.
%\be
%E_{tot} \sim {B_0^2 \over 8\pi} (4 \pi/3) R_0^3 {R_0 \over R_G}
%\ee  
Ref. 
\cite{1979ApJ...227.1013H,2002A&A...384..899M} considered radiation from accelerated changes in the magnetic moment during collapse, producing energy $E\sim B_0^2 (R_G R_{\rm NS})^{3/2}$ (somewhat smaller than the energy of the \Bf\ before the collapse). Along the similar lines,  Ref. \cite{2003ApJ...585..930B}  employed GRMHD simulations and  followed a collapse of a non-rotating \NS\ into the \BH. 

 In our view the main limitation of these models  is that the external medium   was treated as a  vacuum.  Electrodynamically, vacuum is a highly resistive mediums, with the impedance of the order of $4 \pi/c= 477\, \Omega$. As a result, nothing prevents \Bfs\ from becoming disconnected from the star and escaping  to infinity. 
We expect that  the \Bf\  dynamics would be drastically different if the external \ms\ were treated as a highly conducing medium. This is a common consequence in relativistic astrophysical sources, since ample supply of plasma is available through vacuum breakdown.    For example, investigating   the dynamics of the \Bf\ in  the simulations in Ref.   \cite{2003ApJ...585..930B}  \citep[see also][]{2010PhRvD..81l4023L} shows that during the collapse the magnetic field becomes effectively disconnected from the star, at distances somewhat  larger than the \Sc\ radius.
% It is likely that what is observed are the effects of the numerical resistivity.  
If the outside medium were treated as highly conducting plasma, such processes would be prohibited. The importance of resistive effects in the \ms\ was stressed  early on in the original paper by \cite{GinzburgOzernoi}, who point out that   "for spherically-symmetric collapse there is no energy released to the outside at all."

Magnetic field may still escape to infinity if the collapse naturally creates conditions favorable for reconnection, \eg\ by forming narrow current sheets or leading to the overall breakdown of fluid approximation by creating regions where \Ef\ exceed \Bf\ (the latter regions are naturally created both in pulsar \mss\ \citep{2003ApJ...598..446U}
and near \BHs\ moving through the  external \Bf\  \citep{2011PhRvD..83f4001L}). 
In this paper we address a  question  "does the collapse of rotating magnetized \NS\ naturally creates condition for efficient reconnection of magnetic field lines well before the foot points cross the horizon?" We argue that this does not happen. 

The plan of the paper is the following. In \S \ref{precur} we discuss possible types of  precursor emission in NS-NS, NS-BH and BH-BH mergers. In \S  \ref{coll} we make estimates of the  classical (non-GR-modified) pulsar-like  prompt \EM\ power during collapse. In the main \S  \ref{GS-time} we find exact solutions for the structure of collapsing \mss. Based on this solution we argue that  as long as the resulting \BH\ can produce plasma and currents by vacuum breakdown, it  may produce \EM\ much longer that the collapse time.

\section{Precursor emission in mergers}
\label{precur}

For  merging compact objects (NS-NS, BH-NS, BH-BH) a number of mechanism can generate precursor or afterglow emission. 
In case of merging  \NSs,  one expects an \EM\ precursor due to effective 'friction'  of the \NSs' \mss\ \citep{1965JGR....70.3131D,1969ApJ...156...59G,2001MNRAS.322..695H}. Qualitatively, a \NS\ moving through \Bf\ generates an inductive potential drop, which induced real charges on the surface, which in turn  produce a component of the \Ef\ along the \Bf\ and  electric currents. The estimate of  the corresponding power is 
\be
L_ U \approx B_{\rm NS}^2 R_{\rm NS}^2 \beta^2 c = B_{\rm NS}^2  G M  {  R_{\rm NS}^8 \over R_{\rm orb}^7 c} , 
\label{L1}
\ee
where $B_{\rm NS}$ is the surface \Bf\ of a \NS, $ R_{\rm NS}$ is its initial radius and $M$ is its mass, $\beta = v/c$ is the dimensionless velocity of a \NS.
The last equality  in Eq. (\ref{L1}) assumes a Keplerian orbit with radius $R_{\rm orb} $. The estimate (\ref{L1}) can be derived by calculating the potential drop across the \NS, $\Delta \Phi \approx \beta B_{\rm NS} R_{\rm NS}$ and assuming the resistance of the resulting electric circuit to be close to the vacuum inductance $\sim 4\pi/c$.

 Just before the contact, the unipolar power (\ref{L1}) 
 is
\be
L_{U,max}  = 6 \times  10^{45} {\rm erg s} ^{-1} \left( { B _{\rm NS} \over 10^{12} {\rm G}} \right)^2
\ee
for $M_{\rm NS} = 1.4 M_\odot$ and $R_{\rm NS}= 10$ km.

The total \EM\ energy produced by the unipolar induction mechanism can be found by integrating power (\ref{L1}) with the radius evolving due to radiation of gravitational waves, $ R=R_{\rm LC} \left ( 1 - ( G M)^3 t / (c^5 R_{\rm LC}^4) \right)^{1/4}$ and \Bf\ scaling as $B= B_{\rm NS} ( R_{NS}/ R)^3$  (the model becomes applicable when the \mss\ of the \NSs\ touch at the light cylinder distance $ R_{\rm LC}$; at earlier time the interaction is through winds and scales as a sum of the spin-down powers of the \NSs), see \cite{2001MNRAS.322..695H}, 
\be
E_{\rm tot, U}=  \int_ {t(R_{\rm LC})}  ^{ t(  R_{\rm NS })} L_{U, GR} dt \approx  B_{\rm NS}^2 R_{\rm NS}^3 \left( { R_{\rm NS} \over R_{G}} \right)^2 = 
3 \times 10^{43} {\rm erg} \left( { B _{\rm NS} \over 10^{12} {\rm G}} \right)^2
\label{Ett}
\ee

In addition, there is a purely general relativistic effect, when the motion of the compact object across \Bf\  {\it in vacuum}  generates parallel \Ef,  which in turn leads to generation of plasma and the production of \EM\  outflows with power  \citep{2011PhRvD..83f4001L} 
\be 
L _{U, GR}= { (G M)^2 B_0^2 \beta^2 \over  c^3 } =  { (G M)^3 B_0^2 \over  c^5 R_{\rm orb} } 
\label{LU}
\ee 
\citep[see also][]{2010PhRvD..82d4045P,2011arXiv1101.1969M}.
This type of interaction is important for  BH-NS and  BH-BH mergers, in which case there are no real  induced  charges to produce the parallel \Ef,  the parallel \Ef\ is a pure vacuum effect,  resulting  from the  curvature of  the space-time. This power is smaller than for NS-NS coalescence by a factor $  \left( { R_{\rm G} / R_{\rm NS}} \right)^2$, where $R_G =  2  G M /c^2$ is the \Sc\ radius. 

Qualitatively, the power (\ref{LU}) can be estimated from the potential drop across the \Sc\ horizon. There is an important difference between NS-NS  and BH-NS \EM\ interaction, though: in case of the NS-NS system,  the parallel \Ef\ is produced by real surface charges \citep{GoldreichJulian}, while in case of the  \BHs\ the parallel \Ef\ is a pure vacuum effect,  resulting  from the  curvature of  the space-time   \citep{2011PhRvD..83f4001L}.

 For NS-BH system  just before the contact, the general relativistic unipolar power $L_{U, GR}$ is
\be
L_{U, GR}  = 3 \times  10^{44} {\rm erg s} ^{-1} \left( { B _{\rm NS} \over 10^{12} {\rm G}} \right)^2
\label{GGG}
\ee
The total emitted energy
is \be
E_{\rm tot, U}=  \int_ {t(R_{\rm LC})}  ^{ t(  R_{\rm NS })} L_{U, GR} dt \approx  B_0^2 R_{\rm NS}^3 =  10^{42} {\rm erg} \left( { B _{\rm NS} \over 10^{12} {\rm G}} \right)^2
\label{Ett1}
\ee
(Relations (\ref{GGG}-\ref{Ett1}) assume equal masses of the merging objects; it is straightforward to generalize them to unequal masses.)
Thus, the total energy dissipated via the general-relativistic  unipolar induction mechanism is of the order of the magnetic energy of the \NS. Note, that the energy is taken {\it from the linear motion of the \NSs, and  not from the energy of the \Bf. }

 In addition, a more involved \EM\ signatures are expected due to the perturbations that the merging \BHs\ induce in the possible  surrounding gas \citep{2002ApJ...567L...9A,2010ApJ...709..774K,2005ApJ...622L..93M,2010ApJ...711L..89V,2010MNRAS.407.2007C,2011MNRAS.412...75S}.

\section{Pulsar-like  prompt \EM\ power during collapse}
\label{coll}
In this section we discuss pulsar-like  \EM\ power during the  prompt stage of the \NS\ collapse, treating the collapse approximately, in a classical regime up to the \Sc\ radius    $  R_{G} $.  
As the \NS\ collapses, it spins up, $\Omega \propto {R}^{-2}$, \Bf\ increases due to flux conservation, $B  \propto R^{-2}$, while the  radius decreases. Let us first discuss how \EM\ power evolves during the prompt stage of the collapse, neglecting, for the time being, the effects of General Relativity. 

 If the dipole spin-down formula remains valid, the  dipolar \EM\ power increases according to 
\be
L_{\rm d}  \propto B_s^2 R_s^6 \Omega^4 = L_{\rm NS} \left(  { R_{\rm NS} \over R} \right)^6
\label{L}
\ee
where $L_{\rm NS}  \approx B_{\rm NS}^2 R_{\rm NS}^6 \Omega_{\rm NS}^4/ c^3$ is the standard pulsar dipolar spin-down. (Note that the magnetic moment $\propto BR^3 \propto R$ decreases during the collapse.) 
In case of a  free-fall of the \NS\  surface, $ R_s  =R_{\rm NS} (1 - t/t_c)^{2/3}$, where $t_c = (2/3) R_{\rm NS}^{3/2} /\sqrt{ G M } $, resulting in luminosity evolution
 \be
L ={ L_ {\rm NS}  \over (1-t/t_c)^4}
\ee
Limiting the collapse to the fall time down to the \Sc\ radius, $t_f = (2\sqrt{2} /3) ( R_{\rm NS}^{3/2} -  R_G^{3/2}) /( c \sqrt{R_G})$, the total released energy is  relatively small.
\be
E_{tot} ={ 2 \sqrt{2} \over 9} { ( R_{\rm NS}^{9/2} - R_G ^{9/2}) R_{\rm NS}^{3/2} \over c R_G^5} L_{\rm NS} \approx  L_{\rm NS} R_{\rm NS}/c
\ee

The pulsar-like  luminosity of the collapsing \NS\ may be a bit larger than given by Eq. (\ref{L}). 
Under the ideal MHD condition, the \Bf\ is frozen into plasma. Thus, for  field lines penetrating the star,   the  angular velocity  of the field lines is locally equal to the angular velocity of the foot-point.  
The collapse is expected to produce strong 
shearing of the  \Bf\ lines' foot-points.  As a result, large scale currents will be launched into the \ms, increasing the spin-down power.  Increased currents will tend to inflate the \ms, resulting in an increased magnetic flux through the light cylinder and higher spin-down luminosity \citep{tlk}. 
As the upper limit, we can use the spin-down power of the split monopole solution,
 \be
 L_{\rm m } = {2 \over 3} {  ( B_s R_s^2)^2 \Omega_s^2 \over c} \approx \left( { c\over R_{\rm NS}  \Omega _{\rm NS} } \right)^2 \left(  { R_{\rm NS} \over R_s} \right)^4  L_{\rm NS} =\left( { c\over R_{\rm NS}  \Omega _{\rm NS} } \right)^2 { L_{\rm NS}  \over (1-t/t_c)^{8/3}}
  \label{1212}
 \ee
Larger currents in the \ms\  lead  to the increase of power, but for  the collapsing \NS\ the power increases slower with the decreasing radius. As a result,
the total energy released during the collapse time $t_c$  remains fairly small
\be
E_{tot} \approx  \left( { c\over R_{\rm NS}  \Omega _{\rm NS} } \right)^2    {L_0 \over \Omega_0} t_c 
\label{Etot}
\ee
Since the collapse time is short  and $R_{\rm NS}$ not much larger than $R_G$,  the increase in power is mild at best, while the total released energy is small.
 
 \section{Magnetospheres of collapsing \NSs}
 \label{GS-time}

\subsection{Direct emission of \EM\ waves during collapse}

As a \NS\ experiences a collapse, the frozen-in \Bf\ evolves with time, generating \Ef\ and a possible \EM\ signal. Historically, the first treatment of the \EM\ fields of a collapsing \NSs\ was done in the quasi-static approach \citep{GinzburgOzernoi}, in which case the \Ef\ follows from the slow evolution of the \Bf. The quasi-static approach was later demonstrated to give the incorrect asymptotic decay of the fields with time \citep{1972PhRvD...5.2439P}. As the \NS\  contracts, the magnetic moment decreases $\propto R_s$.
% in addition, due to the purely GR effect, as the stellar surface approaches the horizon, the magnetic dipole moment approaches zero \citep{1972PhRvD...5.2439P}.
The scaling of the decay of the fields on the BH calculated in Ref.   \citep{1972PhRvD...5.2439P} was confirmed by \cite{2003ApJ...585..930B}, who performed numerical simulations of the \NS\ collapse into BH and saw a predicted power-law decay of the \EM\ fields.

Most of  the power in the  calculations done in Ref.   \cite{2003ApJ...585..930B}  was emitted at times of the order of the collapse time, well before the predicted  asymptotic limit. Overall, the  simulations are dominated by  heavy resistivity effects intrinsic to the vacuum approximation: the disconnection of the \Bf\ field lines from the star typically  (except in the Kerr-Schild coordinates) occurs at the time when the strong compression of the \Bf\ against the horizon and the corresponding effects of the numerical  resistivity becomes important. 

The assumption of a highly conducting exterior changes the overall dynamics of the \EM\ fields.
As we argue below, the high conductivity of the external plasma would prevent  the formation of disconnected magnetic surfaces, formally prohibiting the processes described  in Ref.   \cite{2003ApJ...585..930B}. 

 \subsection{Force-free approximation in General Relativitiy}
 
 There is a broad range  of astrophysical problems where the  magnetic fields
play a  dominant role, controlling the dynamics of the plasma \citep{2002luml.conf..381B}. The prime examples are  pulsar and black hole magnetospheres;
Gamma-Ray bursts, AGN jet may also be magnetically dominated at 
some stage \cite[\eg][]{LyutikovJPh}.   If the magnetic field energy density  dominates over the plasma
 energy density, the  fluid velocity, 
enthalpy density and a pressure become small perturbations to the  magnetic forces. The dynamics then can be described in a force-free approximation   \citep{Gruzinov99}. 
In the non-relativistic  plasma the notion of force-free fields is often related
to the stationary configuration attained asymptotically  by the system
(subject to some boundary conditions and some constraints, \eg conservation 
of helicity). This equilibrium is attained on time scales of the order of the
\Alfven crossing times. In  strongly magnetized 
relativistic plasma the \Alfven speed may become of the order of the 
speed of light $c$, so that the 
 crossing times becomes of the order of the light travel time. But if plasma
is moving relativistically its state is changing on the same time scale. 
This leads to a notion of dynamical force-free fields. 

MHD formulation assumes  (explicitly) that the second Poincare 
electro-magnetic invariant $\E \cdot \B=0$
and (implicitly) that the first  electro-magnetic invariant is positive
$B^2 -E^2 > 0$. This means  that the electro-magnetic stress energy tensor can be
diagonalized and, equivalently,  that there
is a  reference frame where the  electric  field is  equal to
zero, the plasma rest frame.  This assumption is important since we are interested in the limit
when matter contribution to the stress energy tensor goes to zero;
the possibility of diagonalization of the electro-magnetic stress energy tensor
distinguishes the 
force-free plasma  and vacuum electro-magnetic fields,   where such diagonalization
is generally not possible.

The equations of the  force-free electrodynamic can be derived from Maxwell equations and a constraint $\E\cdot \B=0$.  This can be done in a general tensorial notations from the general  relativistic MHD formulation in the limit of negligible inertia \citep{1997PhRvE..56.2181U}. 
This offers an advantage that the system of equations may be set in the form
of conservation laws \cite{2002MNRAS.336..759K}.  A more practically appealing formulation involves the 3+1 splitting of the equations of general relativity \cite{ThornMembrane,1989PhRvD..39.2933Z}.  The  Maxwell equations in the Kerr metric  then take  the form
 \ba && 
 \nabla \cdot \E =  4 \pi \rho
 \nn &&
 \nabla \cdot \B=0
 \nn &&
 \nabla  \times ( \alpha \B) = 4 \pi \alpha {\vec{j}} + D_t \E
 \nn &&
 \nabla   \times ( \alpha \E) = - D_t \B
 \label{maxw1}
 \ea
  where $D_t = \partial _t - {\cal L} _{\vec{\beta}} $ is the total time derivative, including Lie derivative along the velocity of the zero angular momentum observers (ZAMOs),  $ \nabla$ is a covariant derivative with the radial vector ${\bf e} _r = \alpha \partial_r$ and $\alpha = \sqrt{1-2 M/r}$.
Taking  the total time derivative of the constraint $\E \cdot \B =0$ and eliminating  $D_t \E$ and   $D_t \B$ using Maxwell equations, one arrives at the corresponding  Ohm's law in Kerr  metric \citep{2011PhRvD..83f4001L}, generalizing the result of  \cite{Gruzinov99}:
 \be
 {\vec{j}} = {\left( \B \cdot \nabla \times (\alpha \B) -\E \cdot \nabla \times (\alpha \E)  \right)  \B  + \alpha (\nabla \cdot \E) \E \times \B \over 4 \pi \alpha B^2}
\label{GS000}
\ee
Note that this expression does not contain the shift  function $\vec{\beta}$.  

The generic limitation of the force-free formulation of MHD is that  the evolution of the
electromagnetic
field leads, under certain conditions,  to the formation of regions with $E>B$ \citep[\eg][]{2003ApJ...598..446U}, since there is no
mathematical limitation on $B^2-E^2$ changing a  sign under  a strict
force-free conditions.
 In practice, the particles in these regions  are subject to rapid acceleration
through $\vec E\times\vec B$ drift, following by a formation of pair plasma via various radiative effects and reduction of the \Ef. Thus, regions with $E>B$ are necessarily resistive. This  breaks the ideal assumption and leads to the  slippage of \Bf\ lines with respect to plasma.
In addition, evolution of the  magnetized plasma often  leads to formation of resistive current sheets, with the similar effect on \Bf. If such 
processes were to happen in the \mss\ of the collapsing \NS, this might potentially lead to disconnection of the \Bf\ lines form the star and a \Bf-powered signal.  Below we argue that in case of collapsing \NSs\  this does not happen. 

\section{The  restricted  wave  Grad-Shafranov equations}
\label{PP}
Let us   derive a dynamic  equation that  describes the temporal evolution of the force-free fields in special relativity under the  assumption that the fields remain axially-symmetric. Previously the  equations governing general time-dependent force-free motion 
has been written by \cite{1989ApJ...337...78P,1989ApJ...347..684P}. 

In relativistic plasma the force-free condition is given by the Ohm's law (\ref{GS000}), where in this section we set $\alpha =1$.
 Generally, any   function can be represented as a sum of a gradient and a curl of a vector function. Under the assumption of axial symmetry and zero divergence for \Bf, 
we can express  electric and \Bfs\ as 
\ba && 
  {\bf B} = { \nabla P \times \hat{e_\phi} \over r  \sin \theta } 
 - {2 I \over r  \sin \theta } \hat{e_\phi}
 \nn &&
 {\bf E} = - \nabla \Phi + 
 { \nabla K  \times \hat{e_\phi} \over r  \sin \theta } 
 - {2 L \over r  \sin \theta } \hat{e_\phi}
\ea
where $P$ is the magnetic flux function $P = A_\phi \varpi$, $\varpi =  r\sin \theta$, $A_\phi $, is the electric potential and $K$ and $L$ are some arbitrary functions to be determined, $I$ is the poloidal current through a flux cross-section divided by $2 \pi$. 
The Maxwell equation  $\partial_t {\bf B}= - \nabla \times {\bf E} $
gives
\ba &&
L = \partial_t P /2
\label{DD1}
\\ &&
\partial_t I = { 1 \over 2} \left( 
\partial_r ^2 K  +
{ 1 \over r^2}  \sin \theta \partial_\theta
{ \partial_\theta K \over  \sin \theta } \right) =  { 1 \over 2} \Delta ^\ast K
\nn &&
 \Delta ^\ast  = r^2 \sin ^2 \theta  \nabla \left( { \nabla \over r^2 \sin ^2 \theta} \right)
\label{DD}
\ea
%(The first relation is obvious   by virtue $P = A_\phi \varpi$.)
The ideal condition $\E \cdot \B=0$ implies
\be
2 I \partial_t P =
% - \nabla K \cdot \nabla P + r \sin \theta \nabla \Phi \times \nabla  P \cdot {\bf e}_\phi =
 - 
\left(  \nabla K + r \sin \theta \nabla \Phi \times {\bf e}_\phi  \right)  \cdot  \nabla  P
\label{gg}
\ee
Equations (\ref{DD1})-(\ref{gg}) highlight two separate types of non-stationarity: (i) due to the variations of the current $I(t)$ for a given shape of the flux function (Eq. (\ref{DD})); (ii)  due to the variations of the shape of the flux function  for a given current $I$ (Eq. (\ref{gg})).

% Note, if $K=K(P)$, then for Michel we would have $\partial_t I =0$.

\subsection{Constant shape of flux functions, $\partial_t P=0$, variable current} 
Let us first consider  the case when $\partial_t P=0$. Then Eq. (\ref{gg}) implies that
$
\nabla K _0 + r \sin \theta \nabla \Phi \times {\bf e}_\phi 
$
is orthogonal to $\nabla P$ (and is thus along the poloidal  \Bf).  Above, $K_0$ denotes a particular case when the $P$ is constant in time.  Thus
\ba &&
\nabla K_0  =  - r \sin \theta \nabla \Phi \times {\bf e}_\phi +   { r \sin \theta   \Omega} {  \nabla P \times  {\bf e}_\phi }
\\ \label{34} &&
 \E = -\Omega \nabla P =  - v_{\phi} {\bf e}_\phi  \times \B
 \nn &&
 \partial_ t I =-   { r \sin \theta \over 2}  \nabla P \times  \nabla \Omega \cdot   {\bf e}_\phi = { \varpi^2 \over 2} (\B \cdot  \nabla \Omega)
 \label{dotI}
 \ea
where $\Omega$ is an arbitrary function, which can  be identified with the angular velocity of the rotation. 

The $\phi$ component of the induction equation then becomes the time-dependent Grad-Shafranov equations for the restricted case when the shape of the flux surfaces remain constant, but the angular velocity $\Omega$  and, thus, the  poloidal current are  time and space-dependent:  
\be
\varpi^2 \nabla \left( {1 - \varpi^2 \Omega^2 \over \varpi^2} \nabla P \right) + { 4   I  (\nabla P \cdot \nabla I) \over (\nabla P)^2} + \varpi^2 \Omega(\nabla P \cdot   \nabla \Omega)=0
\label{GS00}
\ee
 This is a Grad-Shafranov for   axisymmetric force-free structures that rotate with arbitrarily varying angular velocity, but keep the shape of the flux functions constant.

The poloidal components of the induction equation  give
\be
\partial _t \Omega = -  { 2 \nabla P \times  \nabla I  \cdot   {\bf e}_\phi \over \varpi (\nabla P)^2}  = {2 \over   (\nabla P)^2} (\B \cdot  \nabla I) 
\label{dotOmega}
\ee
Note that Eqns (\ref{dotI}) and  (\ref{dotOmega})  involve only poloidal \Bf\, which under assumption $\partial_t P =0$ remain constant in time. 

Eqns (\ref{dotI}) and  (\ref{dotOmega}) can be combined to determine the evolution of  $\Omega$:
\be
\partial^2 _t \Omega  = { \varpi^2 \over (\nabla P)^2}  \left(\B \cdot \nabla (\B  \cdot \nabla \Omega) \right) = { \B \cdot \nabla (\B  \cdot \nabla \Omega) \over B_p^2}
\label{omt}
\ee
where $B_p$ is the poloidal \Bf. Eq. (\ref{omt})  is the generalization of the Ferraro's law of iso-rotation to time-dependent angular velocity.

Eqns (\ref{dotI}, \ref{GS00}, \ref{dotOmega}) constitute a closed system of equations for variables $P, \, I,\, \Omega$ under the assumptions of time-dependent $I$ and $\Omega$ and stationary $P$.   Generally, it is not guaranteed that there is a physically meaningful solution of this system: recall that this system describes a restricted motion of force-free plasma, when the shape of the flux function remains constant. 

\subsection{Variable  shape of flux functions}

By virtue of (\ref{gg}) and (\ref{34}) variable shapes of the flux functions can be described by the addition to $\nabla K_0$ of a term proportional to  $\nabla P$,
$K =K_0 + F(P)$. 

Let us first consider $K=F(P)$ separately, neglecting the cross-terms in \Ef.
The $\E \cdot \B=0$ gives
\be
\nabla F \cdot \nabla P =  { 2 I  \partial_t P }
\ee
or, since $F = F(P)$, 
\be
F' (\nabla P)^2=  { 2 I  \partial_t P }
\label{ffg}
\ee

The Maxwell equation  $\partial_t {\bf B}= - \nabla \times {\bf E} $
gives
\be
\partial_t I = {1\over 2} \Delta^\ast F =   {1\over 2}  \left( F'  \Delta^\ast P + (\nabla P)^2 F^{\prime \prime} \right)
\label{dd}
\ee
The $\phi$ component of the induction equation then gives the Grad-Shafranov eq.
\be
 \Delta^\ast P - \partial_t^2 P +  { 4   I  (\nabla P \cdot \nabla I) \over (\nabla P)^2}  - 2 \partial_ t \left( { I^2 \partial_ t  P \over (\nabla P)^2} \right)=0
 \ee
 This is a wave (hyperbolic)  Grad-Shafranov for non-rotating  axisymmetric force-free structures that evolve with time.
The current $I$ here is determined from Eqns. (\ref{ffg})-(\ref{dd}).

The wave Grad-Shafranov equation can be written in a general case, when both current and the flux function evolve with time (Appendix \ref{waveGS}), but it's overly complicated form makes it not useful for practical purposes. 

\subsection{Time-dependent Michel's  split-monopole solution in flat space} 

Both in the  case of accretion induced collapse and for NS-NS mergers, right before the final plunge the NS is expected to rotate with a spin close to break-up limit of $\sim 1 $ msec. As a result, the light cylinder is located close to the NS surface. The theory of pulsar \mss\ predicts that outside the light cylinder the \Bf\ structure resembles the split monopole structure \citep{Michel73}. This is confirmed by direct numerical simulations \citep{Spitkovsky06}. 

In \S \ref{PP} we derived hyperbolic wave Grad-Shafranov equation, describing time-dependent force-free \EM\  fields. It may be verified directly, that  {\it 
 the Michel's  monopole solution  for rotating force-free \ms\ \citep{Michel73} is valid for time-dependent angular velocity $\Omega$, surface \Bf\ $B_s$ and  \NS\ radius $R_s$}. 
For monopole field,  Eq. (\ref{omt}) gives a radially propagating fast wave
\ba && 
\partial^2 _t \Omega  = \partial^2 _r \Omega
\nn && 
\Omega = \Omega(r \pm t)
\ea
The flux  conservations requires $B_s R_s^2 = $const$= B_{\rm NS}  R_{\rm NS}  ^2$.  Then the Grad-Shafranov equation (\ref{GS00}) has a 
 slit-monopole-type solution for \EM\ fields of  the collapsing \NS:
  \ba &&
  B_r = \left( {R_{s} \over r} \right)^2 B_s, \, B_\phi = - { R_{s}^2 \Omega \sin \theta \over r} B_s, \,  E_\theta = B_\phi
  \nn &&
j_r =  -2  \left( {R_{s} \over r} \right)^2  \cos  \theta \Omega B_s 
\nn && 
P= (1-\cos\theta) B_s  R_{s}^2
\nn &&
\Phi= - P \Omega
\nn  &&
I =-{  P (P-2 B_s  R_{s}^2)  \Omega \over 2 B_s  R_{s}^2} = {1\over 2} B_s  R_{s}^2 \Omega \sin ^2\theta
\label{FFF}
\ea
where $P$ is the flux function, and $\Phi$ is the electric potential and $\Omega = \Omega(r-t)$. It may be verified directly that 
Eq. (\ref{dotI}) is satisfied.

Thus, we found exact solutions for time-dependent non-linear relativistic force-free configurations.
Though the configuration  is non-stationary (there is a time-dependent propagating
wave),  the form
of the flux surfaces remains constant. 

\section{Electrodynamics of \NS\ collapse}

 \subsection{Force-free collapse in \Sc\ metric}

Next we apply the solutions obtained in the previous section to the electrodynamics of \NS\ collapse taking into account general relativistic effects. The split monopole solution may be a good approximation for several reasons. First, 
 the collapse is  likely to induce strong shear of the surface foot-points. As a result,  strong electric current will be launched in the \ms\ strongly distorting it. Highly twisted \Bf\ lines will tent to open up to infinity, so that the \ms\ will resemble a monopolar solution at each moment  corresponding to the changing angular velocity of the surface foot-points.  For a general case of strongly sheared foot-points,  a time-depended angular velocity will break a force-balance. Still, we expect that the overall dynamical behavior will  be similar to the time-dependent  Michel's solution. 
 
\begin{figure}[h!]
\includegraphics[width=\linewidth]{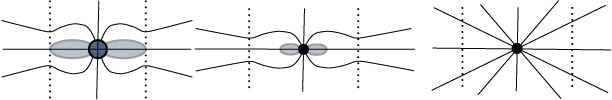}
\caption{Cartoon of the structure of magnetic fields around  a collapsing  rotating \NS. Initially, left panel, the \Bf\ is that of an isolated pulsar, with a set of field lines closing within the light cylinder (dashed vertical lines).  Immediately after the collapse, central panel, the structure is similar. The closed field lines are absorbed by the \BH, while the open field lines remain attached to the \BH; the system relaxes to the monopole structure (right panel).}
\label{NS-Collapse-picture}
\end{figure}

 Second, as we argue below, the open field lines cannot slip off the horizon, while the closed field lines will quickly be absorbed by the \BH. Thus, the \ms\ of the \BH\ will naturally evolve towards the split monopole solution, Fig. \ref{NS-Collapse-picture}.
 Finally,  in a more restricted sense,  the fully analytically solvable dynamics of the monopolar magnetosphere collapse can be used to estimate the physical effects occurring on the   open field lines.

   The stationary Michel's solution has been generalized to \Sc\ metric \citep{BlandfordZnajek} (BZ below).   Extending the time-dependent solution (\ref{FFF})  to the general relativistic case by the principle of minimal coupling (or the  convention "comma becomes a semi-colon"), the Michel's solution (\ref{FFF}) remains valid for arbitrary $\Omega(r_{\rm fast}-t)$ in General relativity. The argument of $\Omega$ should be evaluated at the position of a radially propagating fast mode in the \Sc\  metric with  $dr_{\rm fast}/dt = \alpha^2$, 
\be
\Omega \equiv \Omega\left( r-t + r ( 1-\alpha^2) \ln ( r \alpha^2) \right)
\label{OO}
\ee
The Michel solution in GR has the same flux function as in the flat space (see Eq. (\ref{FFF})), the poloidal \Bf\ is derived from $\Phi$ using a covariant derivative, while toroidal \Bf\ and poloidal \Ef\ change according to  $B_\phi \rightarrow B_\phi /\alpha$ and $E_\theta \rightarrow E_\theta /\alpha$.
   Thus, the exact  non-linear general relativistic time-dependent force-free fields corresponding to the arbitrary solid-body rotation are
     \ba &&
  B_r = \left( {R_{s} \over r} \right)^2 B_s, 
  \nn && B_\phi = - { R_{s}^2 \Omega \sin \theta \over \alpha r} B_s, \,  E_\theta = B_\phi
  \nn &&
j_r =  -2  \left( {R_{s} \over r} \right)^2 {  \cos  \theta \Omega B_s  \over \alpha}
\label{FFF1}
\ea
with $\Omega$ given by Eq. (\ref{OO}). 
It may be verified by direct calculations that fields (\ref{FFF1}) satisfy the Maxwell equations (\ref{maxw1}) with the Ohm's law (\ref{GS000}). 

As the surface of the  \NS\ approaches the \BH\ horizon, $R_s \rightarrow R_G$, $B_s \rightarrow (R_{\rm NS}/R_G)^2 B_{\rm NS} $, while its angular velocity approaches a finite limit which we estimate next. Let initially the \NS\ rotate with angular velocity $\Omega_{\rm NS} $.
The moment of inertia of a neutron star can be written
\be
I _{\rm NS}= (2/5) \chi M_{\rm NS} R_{\rm NS}^2
\ee
where $\chi \sim 0.1 - 0.5$ is an equation of state dependent variable that describes how centrally condensed the star is \citep{1994ApJ...424..823C}.  The spin angular momentum is thus
\be
S = (2/5) \chi M_{\rm NS} R_{\rm NS}^2 \Omega_{\rm NS}
\ee
where $P_{\rm NS}$ is the initial spin period. 
The dimensionless Kerr parameter is  then 
\be
a = (2/5)\chi  (c/G)  { R_{\rm NS}^2  \Omega_{\rm NS} \over M_{\rm NS}}
= 0.04 \chi_{-1}  P_{\rm NS,-3}^{-1}
\label{a}
\ee
where $ P_{\rm NS,-3} = P_{\rm NS}/1 {\rm msec}$.  For merging \NSs\ the Kerr parameter is expected to be much higher. 

 For a collapsing star, the time dilation near the horizon and the frame-dragging of the horizon lead to the  "horizon locking" condition: objects are dragged into corotation with the hole's event horizon, which has a frequency associated with it of
\be
\Omega_H =  a {c^3 \over 2 G r_H} \approx  {\chi \over 5} { c^4R_{\rm NS}^2 \over (G M_{\rm NS})^2}   \Omega_{\rm NS}  =  2.9  \times  10^3 {\rm rad s^{-1}}  \chi_{-1} P_{\rm NS,-3}^{-1}
\ee
 where  $r_H = (1 + \sqrt{1 - a^2})  G M/c^2 \approx R_G$ is the coordinate radius of the horizon of the Kerr \BH.  (Note that for the chosen parameters the final spin is smaller than the initial spin, $\Omega_H /\Omega_{\rm NS} =  0.46\chi_{-1} $,   due to the assumption of highly centrally concentrated initial  mass distribution, $\chi \ll 1$.) 
 
 The \EM\ power produced by the Michel's rotator is then (see Eq. (\ref{1212}))
 \be
 L= {2 \over 3} {  ( B_s R_s^2)^2 \Omega_H^2 \over c}= {2 \over 75}\chi^2 { c^7 B_{\rm NS}^2 R_{\rm NS}^8 \Omega_H^2 \over ( G M_{\rm NS})^4} =
 2 \times 10^{44} \, {\rm erg s}^{-1}\, \chi_{-1}^{2}\,  B_{\rm NS,12}^{-2} \, P_{\rm NS,-3}^{-2}  
 \label{111}
 \ee
 It will lead to the \BH\ spin-down on a time scale
 \be
 \tau =6 { G^2 M_{\rm NS}^3 \over c^{3} B_{\rm NS}^2 R_{\rm NS}^4 }= 2 \times 10^7  \, {\rm sec} \,  B_{\rm NS,12}^{-2}
 \label{tauSD}
\ee
(Michel solution corresponds to the spin-down index of $n=1$, so that the spin evolution is described by a decaying exponential.) 
It is unlikely, though, that the assumptions of the model will be applicable for such a long time, see below. 
 
 In addition, the \NS\ with dipolar \Bf\  has a net charge  
$Q= (1/3)  B_{\rm NS} r_{\rm NS}^3 \Omega_{\rm NS}/c$.
As long as the assumption of  the model are satisfied (that the \BH\ produces a wind, see below), this charge is not canceled by the  electrostatic  attraction of charges from the  surrounding medium. 
Thus, the \BH\ settles to the  Kerr-Newman solution. The corresponding Newman parameter is small
\be 
 b= { \sqrt{ Q^2 G /c^4} \over R_H}={ Q \over 2 \sqrt{G} M _{\rm NS}} ={ B_{\rm NS} R_{\rm NS}^3 \Omega_{\rm NS} \over 6 c  M_{\rm NS}} = 
4\times  10^{-8}  B_{\rm NS,12} \, P_{\rm NS,-3}^{-1}
 \label{b}
 \ee
 As we argued above, the closed \Bf\ lines will be quickly absorbed by the \BH, so that the \ms\ will settle to the monopolar \Bf\ structure with no electric charge,   Fig. \ref{NS-Collapse-picture}.
 
 \subsection{How a \NS\ collapse proceeds}
 
 To summarize the above discussion,  first, the space-time  of the  collapsing \NS\ temporarily  passes through the Kerr-Newman  solution with parameters given by (\ref{a}),  (\ref{b}); quickly the electric charge is lost due to the absorption of the closed field lines. (We stress that the loss of the electric charge is driven by the internal electrodynamics and not by the attraction of charges from the surrounding medium.) 
 Second, and most importantly,   we have demonstrated that collapsing \NS\ does not produce any narrow current structures or other dissipative/resistive structure
 that could have became dissipative and "released" the overlaying \Bf\ to the infinity: the field always remain connected to the surface of the star. 
 
 The fate of the \Bf\ lines connected to the surface of the star then depends on whether it is a closed magnetic field line, or the one open to infinity. For closed loops, both  footpoints are dragged toward the horizon and eventually absorbed by the \BH. On the other hand, the open \Bf\ lines remain open and connected to the hole, without "sliding off the \BH",  as long as the assumptions of the model remain satisfied. 
  Thus, for \BH\ surrounded by  highly conducting plasma  {\it the open  \Bf\  lines never become disconnected from the \BH}. 
 As a result, the \EM\ power emitted by the \BH\ may  continue for times much longer than the immediate collapse time. 
 
 The key difference here from the conventional  BZ mechanism is that in the latter case the  \Bf\ is assumed to be  produced by the currents in {\it externally supplied  accretion disk}, while here  the  \Bf\ 
 is produced by the currents generated by the \BH\ itself. Also note, that 
 this result does not violate the "no hair" theorem \citep[\eg][]{MTW}, which  assumes that the outside is  vacuum.  In our case the outside medium is assumed to be high conducting plasma all the way down to the \BH\ horizon. Under this assumption  the \Bf\ lines cannot disconnect from the \BH.  
 
 There is a  natural limit of applicability of  the present model. The electric currents that support the \Bf\ on the \BH\ are assumed to be self-produced by the \BH\ via  the vacuum breakdown, and not supplied by the external current, like in the BZ case. Vacuum breakdown requires a sufficiently high electric potential. As the \BH\ spins down, the potential available for particle acceleration decreases. After some time, the \BH\ will not be able to break vacuum. It would cross a death-line (using pulsar terminology) after which  moment no particles are produced anymore, the outside becomes vacuum, and by the no hair theorem the \BH\ will lose its \BH. Also, starting this moment the \BH\ will be able to attract charges of the opposite sign, canceling the internal charge.
 
In fact, a somewhat different scenario is likely to play out.   Our experience with pulsars indicate that the plasma production in the \ms\ is a highly non-stationary process. If there is an interruption in the plasma production  for  sufficiently long time, the  magnetic field will able to slide off the \BH, shutting down the \EM\ power forever.
 
   \section{On the nature of short Gamma Ray Bursts}

 The above results further highlight possible difficulties with the progenitors of short GRBs being the merging \NSs\ \citep{Lyutikov:2009}.   On the one hand,  numerical simulations indicate that the active stage of NS-NS coalescence typically takes 10-100 msec. Only  small amount, $\leq 0.1 M_\odot$  of material may be ejected during the merger and accretes on time-scales of 1-10 secs, depending on the assumed $\alpha$ parameter of the resulting  disk \cite[\eg][]{Kiuchi, LiuNSNS,Faber}.   Thus, there is not enough baryonic matter left outside the BH to power a short GRB. Any energetically dominant  activity on much longer time scales  contradicts the NS-NS coalescence paradigm  for short GRBs.  This seem to contradict observations that  some short GRBs have   long extended  $X$-ray tails observed over time scales of tens to hundreds of seconds.  The tail fluence can   dominates over the primary burst \citep[by a factor of 30 as in GRB080503,][]{Perley}. In addition,  powerful flares appear late in the afterglows of both short and long GRBs (\eg  in case of GRB050724 there is a powerful flare at $10^5$ sec). In the standard forward shock model of afterglows this requires that at the end of the activity, lasting 10-100 msec,  the source releases  more energy than during the prompt emission  in a form of  low $\Gamma$ shells, which collides with the  forward shock  after $\sim 10^6$ dynamical times, a highly fine-tuned  scenario.

On the other hand, the expected \EM\  powers estimated in the present paper are fairly low for all the discussed processes.  Since the 
 above results are based on the analytical Michel-type structure  of the \BH\ \mss, which for a given surface \Bf\ and the spin  has the largest amount of open \Bf\ lines and the largest  \EM\ power,   the numerical estimates above can be considered as upper limits.

  The only exception to the above could be that an efficient magnetic dynamo mechanism operates either  during \NS\ merger (for short GRBs) or during a core collapse of a massive star (for long GRBs),  resulting in a formation of a millisecond magnetar-type object with \Bf\ reaching $10^{14}$ G \citep{Usov92}. Since, as we argue, the \BH\ can retain its \Bf\ for a long period of time, the spindown  time scale (\ref{tauSD}) may become sufficiently short, hundreds to thousands of seconds, so that the \Bf\ can electromagnetically  extract a large fraction of the total rotation energy of the \BH
\be
E_{\rm tot} = {1\over 2} I _{\rm NS}\Omega_{NS}^2 =  2 \times 10^{51} {\rm erg} \, \chi_{-1} \, P_{\rm NS,-3}^{-2}. 
\ee

The fact that the \EM\ extraction of the rotational energy of the \BH\ can operate both in long and shot GRBs may explain a surprising observation  that early afterglows of long and short GRBs  look surprisingly similar, forming a continuous sequence, \eg, in relative intensity of X-ray afterglows as a function of prompt energy \citep{Nysewander}.
 This is surprising in a forward shock model:  the properties of the forward shock do depend on the external density, while the prompt emission is independent of it.  The difference between circumburst media densities in Longs (happening in star forming regions) and Short
(happening in low density galactic or even extragalactic medium) is many orders of magnitude.  
In defense of the forward shock model, one might argue that afterglow dynamics depends on $E_{ejecta} /n$, both of which are orders of magnitude smaller for short GRBs if compared with long GRBs. 
Yet afterglows are very similar and, most importantly, form a continuous sequence

We suggest that the similarity of  the early afterglows  properties of long and short GRBs, ,at times $\leq 10^5 $ sec, can be related to the fact that in both cases a spinning  \BH\ can retains \Bf\ for sufficiently long time to power the prompt and early afterglow emission via internal dissipation in the wind \citep{Lyutikov:2009}.

  \section{Discusion}
  
  In this paper we discuss possible \EM\ signatures of the merging and  collapsing compact objects. At the in-spiraling stage, 
  in case of  NS-NS system,  the peak  Poynting power  is 
 $ L_{U,max}  = 6 \times  10^{45} \, {\rm erg\,  s}^{-1}   \left( { B _{\rm NS} / 10^{12} {\rm G}} \right)^2$, while for BH-NS systems it is an order of magnitude smaller.  Both the peak power and the total energy of the precursor emission are fairly small, see \S \ref{precur}. Only  for magnetar-type \Bfs\ the corresponding emission can be observed at cosmological distances, see \cite{2001MNRAS.322..695H}

   We found  Michel-type solution for  the structure of  time-dependent force-free \mss\ in General relativity. Based on this solution, we argue that  contrary to the previous estimates 
 the direct emission of the \EM\ field, powered by the magnetic energy stored outside of the \NS,  does not produce a considerable \EM\ signal: such process is prohibited by the high conductivity of the surrounding  plasma.

 Most importantly,  as long as the \BH\ is able to produce a highly conducting plasma via the  vacuum breakdown, \Bf\ cannot "slide off" the \BH. As a result, a \BH\ can retain \Bf\ for much  longer  time that is predicted by the "no hair" theorem,  producing an \EM\  power for a long time after the collapse, without a need for an externally supplied \Bf. The "no hair" theorem does not apply here due to the assumed high  
 conductivity of the plasma surrounding the \BH.   (Pulsars produce  plasma and currents  all by themselves, without an external accretion disk.) Since in the force-free limit the structures in the current sheet are flying away with the speed of light \cite[\cf, the corrugated current current sheet solution in Ref.][]{1999A&A...349.1017B}, any \Bf\ reconnection occurring beyond the light cylinder does not affect the global solution. 
The moment the \BH\ fails to produce the plasma (\eg\ due to spontaneous reconnection within the light cylinder), it will quickly lose its \Bf\ and stop producing any \EM\ power. (It takes one malfunction to break the \BH\ \EM\ engine). It will likely to be a random processes, with no typical time-scale, that will terminate the EM emission well before the BH spins down.

 I would like to thank Thomas Baumgarte, Vasily Beskin,  Scott Hughes,  Luis Lehner, Harald Pfeiffer, Eric Poisson,  Stuart Shapiro and Alexander Tchekhovskoy  for discussions. 

\bibliographystyle{apsrev}

  \bibliography{/Users/maxim/Home/Research/BibTex}

\appendix
\section{Hyperbolic  Grad-Shafranov equation}
\label{waveGS}

Generally, we can write 
\ba &&
\nabla K  =  - r \sin \theta \nabla \Phi \times {\bf e}_\phi +   { r \sin \theta   \Omega} {  \nabla P \times  {\bf e}_\phi }+ \nabla F(P)
\nn &&
\E= -\Omega \nabla P  - {\partial_t P \over \varpi} {\bf e}_\phi + \B_p F'
\nn &&
\E \cdot \B=0 \rightarrow F'= {2 I \partial_ t  P \over  (\nabla P)^2} 
\ea

The $\phi$ component of the induction equation gives (the poloidal components are satisfied identically) 
\be
\partial_t I = {1\over 2} \left(  \Delta^\ast F + \varpi (\nabla \Omega \times   \nabla P) \right) = {1\over 2} \left(  \Delta^\ast F + \varpi ^2 (\B \cdot \nabla \Omega) \right)
\ee

The $\phi$ component of the Ampere's law  gives the hyperbolic  wave Grad-Shafranov equation
\ba && 
\varpi^2 \nabla \left( {1 - \varpi^2 \Omega^2 \over \varpi^2} \nabla P \right) - \partial_t^2 P + \left\{ - 4 {(\Delta ^\ast P) I^2    } - 2 { I F^{\prime \prime} \partial_t P  (\nabla P)^2} + 
{ I F^\prime \partial_t (\nabla P)^2 } 
\right.
\nn &&
\left.
{  4 (\nabla P \cdot \nabla I) I  } +{ 4 \Delta^\ast P \varpi^2 I^2 \Omega^2 } - 
{ 2 \varpi I \Omega^2  (\nabla P \times \nabla (\partial_t P/\Omega) ) \cdot {\bf e}_\phi} - 2 {( \Delta^\ast P) I F^\prime \partial_t P } 
\right.
\nn &&
\left.
 + { ( (\nabla P)^2 + 8 I^2) \varpi^2 \Omega (\nabla P \cdot \nabla \Omega) }  +
{ 8 \varpi I^2 \Omega^2 (\nabla P \cdot \nabla ( r \sin\theta))  } \right\} { 1 \over (\nabla P)^2 + 4 I^2} =0
\ea

\end{document}